# Ultrasonic Identification Technique in Recycling of Lithium Ion Batteries

Michiel Postema[1,2,3]
[1]School of Electrical and Information Engineering
University of the Witwatersrand,
Johannesburg, South Africa
Michiel.Postema@wits.ac.za

Satyajit Phadke[2,6]
[2] LE STUDIUM
Loire Valley Institute of Advanced Studies
Orléans, France
satyajit.phadke@univ-tours.fr

Anthony Novell[3]
[3]Inserm Research Unit U1253: iBrain
Faculté de Médecine
Université de Tours
Tours, France
anthony.novell@univ-tours.fr

Rustem Uzbekov[4,5]
[4] Department of Microscopy
Faculté de Médecine
Université de Tours
Tours, France
[5] Faculty of Bioengineering and Informatics
Moscow State University
Moscow, Russia
rustem.uzbekov@univ-tours.fr

Cuthbert Nyamupangedengu[1]
School of Electrical and Information Engineering
University of the Witwatersrand,
Johannesburg, South Africa
cuthbert.nyamupangedengu@wits.ac.za

Mériém Anouti[6]
[6] Physiochimie des Matériaux et Electrolytes pour l'Energie (PCM2E)
Université de Tours
Tours, France
meriem.anouti@univ-tours.fr

Ayache Bouakaz[3]
Inserm Research Unit U1253: iBrain
Faculté de Médecine
Université de Tours
Tours, France
ayache.bouakaz@univ-tours.fr

*Abstract*—The recycling of lithium ion batteries has been mentioned as one of the near-future waste management necessities. In order for recycling to be economically viable, straightforward and cost effective techniques need to be developed to separate the individual materials in a composite electrode. Ultrasonic separation might be such a technique, provided that lithium ion battery microparticles respond predictably to a sound field. Lithium ion battery cathodes contain hydrophobic carbon. Owing to the incompressibility of a solid, the thin gaseous layer surrounding these hydrophobic particles must oscillate asymmetrically, when subjected to ultrasound. Consequently, the harmonic content of the ultrasound signal radiated from hydrophobic microparticles must be higher than that from hydrophilic microparticles with the same size. The question of whether the harmonic signal response generated by physical hydrophobic microparticles present in lithium ion battery cathodes is higher than the harmonic response of other component materials in the cathode is the focus of this paper. The scattering response of cathode materials subjected to 1-MHz ultrasound was measured and compared. The cathode materials C65, PVDF, and NMC respond differently to 1-MHz ultrasound. The superharmonic response of C65 has been attributed to asymmetric oscillations owing to its hydrophobicity. In addition, C65 hydrophobic microparticles might be suitable candidates for harmonic imaging.

## I. Introduction

There are world-wide efforts aimed at achieving energy transition into more environmentally sustainable technologies that have minimal or no carbon footprint. In the electricity sector and ancillary technologies such as those driving the 4[th] Industrial Revolution, the efficiency of energy storage batteries is critical. Consequently, there have been concerted efforts towards continuous improvement of battery technologies. As an example, in the motor vehicle industry, electrical vehicles largely use lithium ion batteries instead of the traditional Lead-acid batteries.

In the context of sustainability, recycling of batteries becomes a standard requirement. However, while Lead-acid batteries are widely recycled, the same cannot be said about lithium ion batteries. Recycling of lithium ion batteries is more challenging due to the wider variety of materials in each cell [1]. Furthermore, the materials are not discrete as in Lead-acid batteries. Despite the challenges, there is ongoing search for viable methods of recycling lithium ion batteries [1]. Various alternatives of recycling lithium ion batteries have been attempted, such as the pyrometallurgical process, the hydrometallurgical process and the direct physical recycling process [2]. In order for recycling to be economically viable, straightforward and cost-effective techniques need to be developed to identify and separate the individual materials in a composite electrode.

Typically, the cathode in a lithium ion battery consists of three components, namely the active material lithium transition metal oxide, conductive carbon particles, and a polymer binder. For construction of the battery electrodes, the three materials are mixed intimately in industrial binders so as to obtain a homogeneous composite cathode material. This cathode material composite is then used for the fabrication of batteries. In order to effectively recycle the materials, the three components need to be individually segregated so that the different materials can be separately processed chemically [3,4].

Ultrasonic separation might be a suitable segregation technique, provided that lithium ion battery microparticles respond predictably to a sound field. The forcing of microparticles by means of ultrasonic manipulation has been studied extensively in the medical field [5,6].

Lithium ion battery cathodes contain hydrophobic carbon, which is used as an additive for enhancing the electrical conductivity of the electrode through a conductive network [7,8]. Owing to the incompressibility of a solid, the thin gaseous layer surrounding these hydrophobic particles must oscillate asymmetrically, *i.e.*, the radial excursion amplitude of the outer gaseous surface is greater during expansion than during contraction, when subjected to ultrasound [9]. Consequently, the harmonic content of the ultrasound signal radiated from hydrophobic microparticles must be higher than that from hydrophilic microparticles of the same size. Therefore, such microparticles might be suitable agents for harmonic imaging. Examples of numerical simulations of the acoustic response from incompressible droplets surrounded by gaseous shells have been presented in [10], indicating a significant higher harmonic content of these so-called antibubbles compared to conventional bubbles without a droplet core.

The question of whether the harmonic signal response generated by physical hydrophobic microparticles present in lithium ion battery cathodes is higher than the harmonic response of other component materials in the cathode is the focus of this paper.

The scattering response of cathode materials subjected to 1-MHz ultrasound was measured and compared. Differences in acoustic response make particles identifiable and therefore easier to separate.

## II. MATERIALS AND METHODS

We prepared four media for this evaluation.

The first material studied was a 1-ml Otec® R0,9% saline (LABORATOIRE AGUETTANT, Lyon, France) emulsion containing 20 mg Li(Ni$_{0.33}$Co$_{0.33}$Mn$_{0.33}$)O$_2$ (NMC) cathode active material (Targray, Kirkland, QC, Canada). The second material studied was a 1-ml saline emulsion containing 11 mg Polyvinylidene Fluoride (PVDF) binder (Targray). The third material studied was a 1-ml saline emulsion containing 12mg TIMCAL SUPER C65 Carbon Black (EQ-Lib-SuperC65) conductive additive (MTI Corporation, Richmond, CA, USA).

In addition, a medium consisting of just saline was similarly studied. The experiments with saline were performed as null experiments.

For illustration of the microparticle sizes, Fig. 1 presents a scanning electron microscope image of a dried mixture of the three materials. In preparation of this image, dry samples were sprinkled onto carbon disks before observation under a Zeiss Ultra plus FEG-SEM scanning electron microscope (Carl Zeiss Microscopy GmbH, Jena, Germany).

The materials to be evaluated were shaken for 60 s in a CapMix™ (3M ESPE, Seefeld, Germany). From these emulsions, 20 µl was pipetted into a Fisherbrand® FB55143 macro cuvette (Fisher Scientific SAS, Illkirch, France) containing 3 ml saline. The cuvette was then placed centrally in a container filled with degassed water.

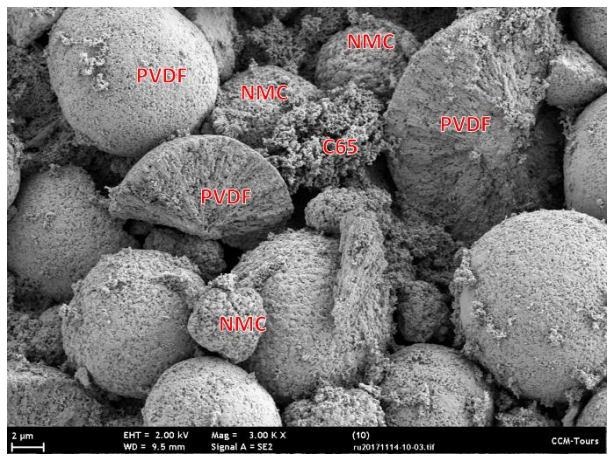

Fig. 1. Scanning electron microscope image of a dried mixture of C$_{65}$, PVDF, and NMC.

The experimental setup used was presented in a previous study [11]. Briefly, as described in [12], an unfocussed transmitting single-element transducer was mounted to one side and a receiving single-element transducer was mounted perpendicularly to the transmitting transducer, as illustrated in Fig. 2. A metal frame was attached to the container so that the media under investigation could be positioned precisely.

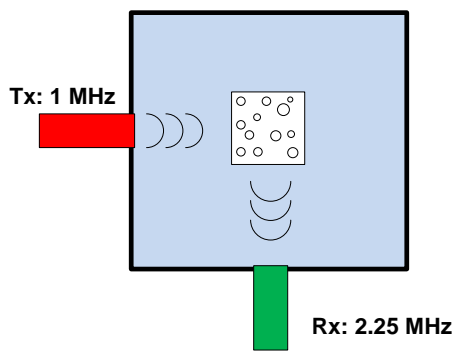

Fig. 2. Top view of the experimental setup [11].

In each experiment, a pulse from a WW5061 50 MS/s waveform generator (Tabor Electronics Ltd., Nesher, Israel) was triggering a 33220A arbitrary function generator (Agilent Technologies, Santa Clara, CA, USA), which generated 20 cycles of a 1-MHz, 100 mV peak-to-peak signal. The signal was attenuated with a 75-A-MFN-03 75-W, 3-dB attenuator (Foshan Yixun Co Ltd, Longjiang, PR China) and subsequently amplified by an AAP-500-0.2-6-D500-W power amplifier (ADECE, Veigne, France). The signal was transmitted with an unfocussed custom 1.0-MHz transmitting single-element transducer (SOFRANEL, Sartrouville, France) with a 13-mm diameter. A custom 2.25-MHz, 60%-bandwidth receiving single-element transducer (SOFRANEL, Sartrouville, France) with a 51-mm diameter and focussed at 55 mm was mounted perpendicularly to the transmitting transducer.

The received signal was amplified by 20 dB using a 5077PR square wave pulser/receiver (Olympus Corporation, Shinjuku, Tokyo, Japan) in receive mode. It was recorded using a TDS 3044B digital oscilloscope (Tektronix, Beaverton, OR, USA). The recorded signal was transferred to a personal computer using a GPIB cable and MATLAB® (The MathWorks, Inc., Natick, MA, USA) software.

The time-delay between transmission and first reception was determined manually before the experiments. The time of first reception was set as origin in the recordings. In each experiment, a response signal with a duration of 100 μs was recorded at a sampling rate of 100 MHz.

Thirty identical experiments were performed for each of the three media containing microparticles, *i.e.*, C65, NMC, and PVDF. Four hundred and fifty identical experiments were performed with saline alone.

Using the Fast Fourier Transform, frequency spectra of the recorded signals were computed in MATLAB®. For each medium studied, the response from the thirty experiments was averaged. Spectral noise was removed with a five-point running smoother. The resulting amplitude spectra were normalised by the amplitude spectrum from saline alone before being presented on a decibel scale.

III. RESULTS AND DISCUSSION

The results from the null experiments with saline are presented in Fig 3.

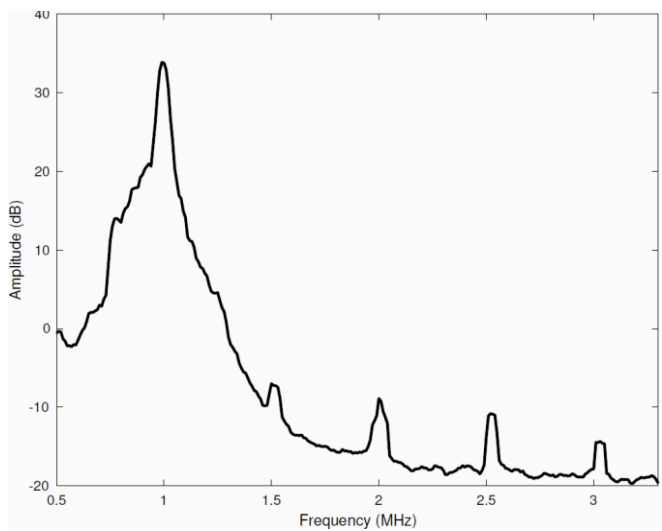

Fig. 3. Fourier spectrum of the acoustic response from saline: amplitude in dB as a function of frequency in MHz.

A wide-band fundamental mode is evident in the response. Not only can higher harmonics be appreciated at 2 MHz and 3 MHz, but also ultraharmonics at 1.5 MHz and 2.5 MHz. This response can be attributed to the geometry of the setup, which allows for multiple reflections from the cuvette surfaces and scattering from its sharp corners.

The results from the experiments with C65 are presented in Fig. 4. The fundamental response around 0 dB indicates that the acoustic response from C65 does not significantly differ from that of the saline medium. However, wide-band higher harmonics at 2 MHz and 3 MHz of more than 20 dB can be appreciated.

The higher harmonics from these hydrophobic microparticles has been attributed to the asymmetry between expansion and contraction predicted from theory.

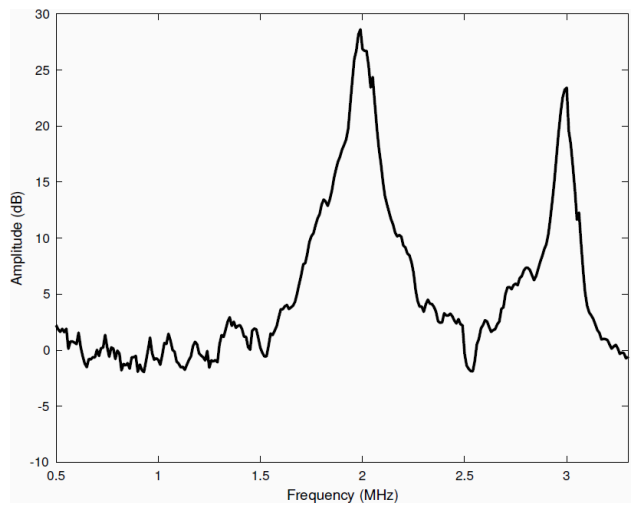

Fig. 4. Fourier spectrum of the acoustic response from C65: amplitude in dB as a function of frequency in MHz.

The results from the experiments with PVDF are presented in Fig. 5. Here, the fundamental response shows two peaks up to 5 dB. Given that the amplitude spectrum has been normalised by the spectrum of the null experiments with saline alone, this means that the fundamental response from PVDF is wide-band. In addition, a narrowband 2-MHz peak is evident.

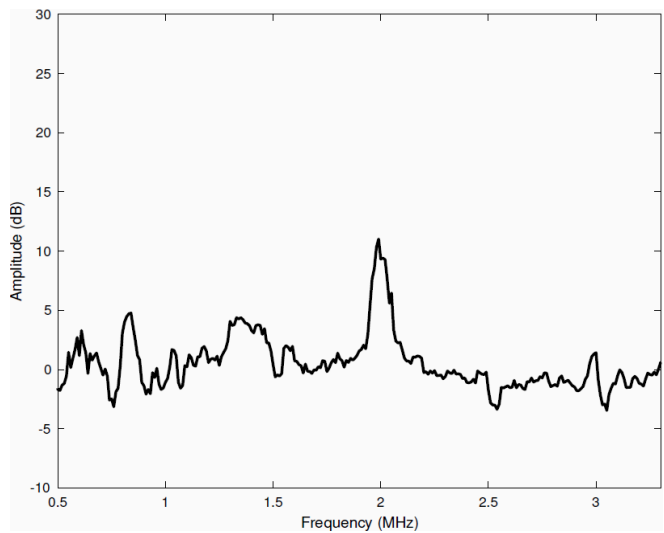

Fig. 5. Fourier spectrum of the acoustic response from PVDF: amplitude in dB as a function of frequency in MHz.

The results from the experiments with NMC are presented in Fig. 6. The response from NMC is below 0 dB, indicating that it acts solely as an acoustic attenuator under the experimental conditions used here. For all experiments, the absence of wide-band noise indicates that the microparticles have remained intact during the experiments.

Owing to the difference in response of the emulsions, the microparticles might be subjected to continuous sound waves, to drive them through liquids at different velocities, causing separation. This will be the purpose of a follow-up study.

These results also indicate the potential of hydrophobic microparticles in harmonic imaging. The frequencies used in these first experiments were chosen with the knowledge of the average size of the microparticles.

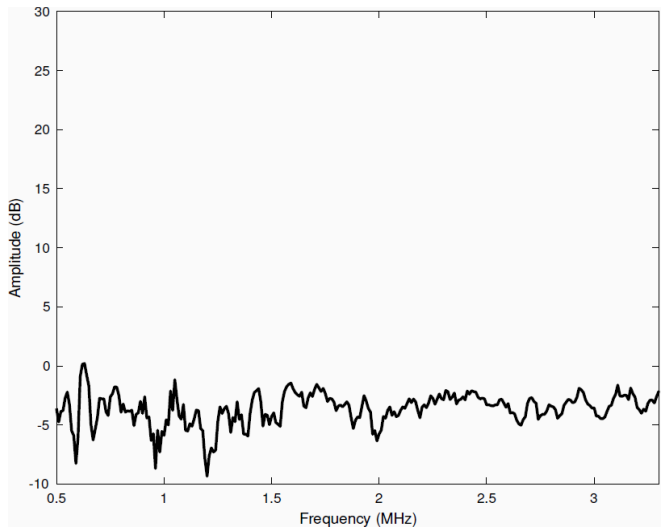

Fig. 6. Fourier spectrum of the acoustic response from NMC: amplitude in dB as a function of frequency in MHz.

IV. CONCLUSION

The cathode materials C65, PVDF, and NMC respond differently to 1-MHz ultrasound. The superharmonic response of C65 has been attributed to asymmetric oscillations owing to its hydrophobicity. In addition, C65 hydrophobic microparticles might be suitable candidates for harmonic imaging.

C65 can be identified based on its characteristic, harmonic acoustic signature. This may have implications for the separation of lithium ion battery components and consequently for the affordable recycling of lithium ion batteries.

ACKNOWLEDGMENT

M.P. and S.P. have received funding from European Union's Horizon 2020 research and innovation programme under Marie Skłodowska-Curie grant agreement No 665790. The scanning electron microscope data were obtained with the aid of the IBiSA Electron Microscopy Facility of the University of Tours and the University Hospital of Tours.